\documentclass[
reprint,
superscriptaddress,
 amsmath,amssymb,
 aps,
floatfix,
]{revtex4-2}

\usepackage{graphicx}
\usepackage{dcolumn}
\usepackage{bm}
\usepackage[version=4]{mhchem}

\usepackage{natbib}
\begin{document}

\title{Spin-orbit Energies in Etch-Confined Superconductor-Semiconductor Nanowires}

\author{J.~D.~S.~Witt}
\email{james.witt@sydney.edu.au}
\affiliation{ARC Centre of Excellence for Engineered Quantum Systems, School of Physics, The University of Sydney, Sydney, NSW 2006, Australia}

\author{G.~C.~Gardner}
\affiliation{Department of Physics and Astronomy, Purdue University, West Lafayette, Indiana, USA}
\affiliation{Microsoft Quantum Purdue, Purdue University, West Lafayette, Indiana, USA}

\author{C.~Thomas} 
\affiliation{Department of Physics and Astronomy, Purdue University, West Lafayette, Indiana, USA}
\affiliation{Microsoft Quantum Purdue, Purdue University, West Lafayette, Indiana, USA}

\author{T.~Lindemann} 
\affiliation{Department of Physics and Astronomy, Purdue University, West Lafayette, Indiana, USA}
\affiliation{Microsoft Quantum Purdue, Purdue University, West Lafayette, Indiana, USA}

\author{S.~Gronin} 
\affiliation{Department of Physics and Astronomy, Purdue University, West Lafayette, Indiana, USA}
\affiliation{Microsoft Quantum Purdue, Purdue University, West Lafayette, Indiana, USA}

\author{M.~J.~Manfra}
\affiliation{Department of Physics and Astronomy, Purdue University, West Lafayette, Indiana, USA}
\affiliation{Microsoft Quantum Purdue, Purdue University, West Lafayette, Indiana, USA}

\author{D.~J.~Reilly}
\email{david.reilly@sydney.edu.au}
\affiliation{ARC Centre of Excellence for Engineered Quantum Systems, School of Physics, The University of Sydney, Sydney, NSW 2006, Australia}
\affiliation{Microsoft Quantum Sydney, The University of Sydney, Sydney, NSW 2006, Australia}

\date{\today}

\begin{abstract}

We report magneto-transport measurements of quasi-1-dimensional (1D) Al-InAs nanowires produced via etching of a hybrid superconductor-semiconductor two-dimensional electron gas (2DEG). Tunnel spectroscopy measurements above the superconducting gap provide a means of identifying the 1D sub-bands associated with the confined 1D region. Fitting the data reveals the strength of the different components of the spin-orbit interaction (SOI), which is of interest for evaluating the suitability of superconductor-semiconductor 2DEGs for realizing Majorana qubits.


\end{abstract}
\maketitle

\section{Introduction}

Hybrid superconductor--semiconductor devices theoretically provide all of the requisite ingredients to generate Majorana quasiparticles. The combination of $s$-wave pairing, strong spin-orbit interaction (SOI), and gate-control over the chemical potential enables the engineering of an effective spinless, $p$-wave, superconducting topological phase. This phase has a multiply degenerate ground state, the edge-states of which are zero-energy modes or Majorana bound states (MBS) -- so-called because they are the quasiparticle analogue of the theoretical Majorana fermion. By virtue of being their own antiparticle, MBS are incapable of coexisting in the same physical space. Thereby, an additional requirement for MBS generation is that the proposed host system possesses an odd number of conduction modes \cite{Lutchyn2018,Frolov2020}.

To-date, experimental efforts to engineer systems capable of hosting MBS have followed three key routes: Vapour-liquid-solid (VLS) nanowires \cite{Mourik2012,Deng2012,Das2012,Churchill2013,Finck2013,Deng2016,Chen2017,Gul2018}; selective-area growth (SAG) molecular beam epitaxy \cite{SAG1,SAG2,SAG3,SAG4,SAG5,SAG6}; and heterostructures supporting 2DEGs, where subtractive etching of material or gate-based depletion is used to define 1D regions \cite{Kjaergaard2016,Shabani2016,Nichele2017,Suominen2017}. VLS nanowires have received much focus, but scale-up of this platform into the complex arrays needed for quantum computing is challenging \cite{VLS1,VLS2}. The alternate approaches perhaps offer more feasible routes for fabricating scalable qubit arrays. The opportunity to suppress disorder by using 2D heterostructures with super-lattice buffers \cite{Manfra,Pauka2020} whilst retaining the flexibility of patterning needed for scaling-up provides further motivation for studying 2DEGs. However, despite their promise, open questions remain concerning the extent to which the favourable properties of 2DEG based systems remain when 1D regions are defined using etching or gate-depletion. 

\begin{figure}[t]
\includegraphics[width=\linewidth]{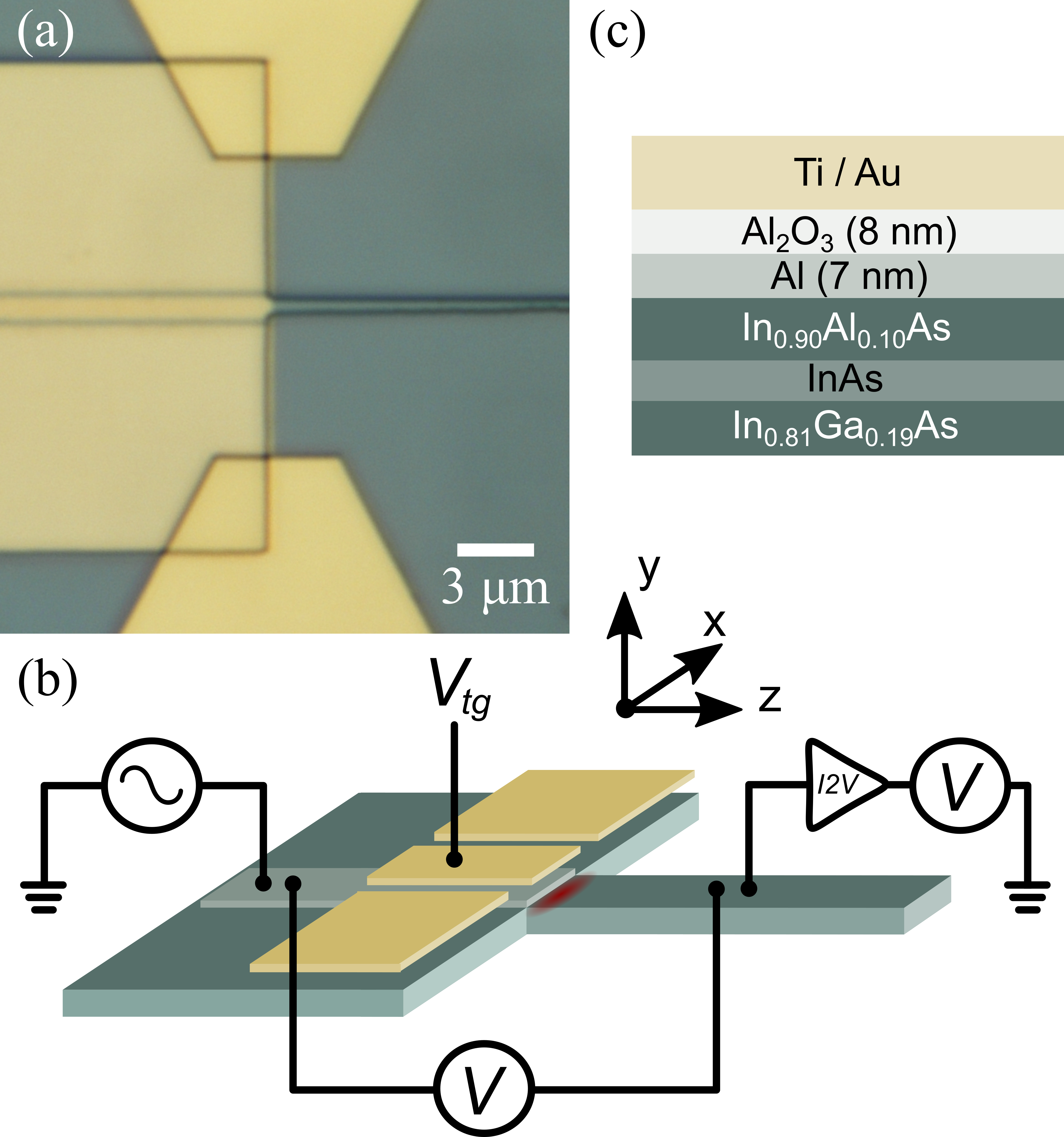}
\caption{\label{fig1} (a) Device micrograph, (b) device schematic indicating the measurement set-up -- 4-point measurements were made by contacting the aluminium strip on the left-hand side of the device and the etched wire (via an Al contact pad) on the right-hand side of the device. The position of the formed tunnel barrier is indicated in red. The magnetic field axes relative to the device are also shown, and (c) the heterostructure stack. The electrostatic gates -- gold colour, the aluminium -- light grey and the InAs mesa -- dark grey throughout.}
\end{figure}

Here, we make use of tunnelling conductance spectroscopy to probe the above-gap regime of a quasi-1D nanowire, formed via the subtractive etching of a superconducting-semiconducting Al-InAs heterostructure. Although tunnel spectroscopy has been widely used for probing the sub-gap structure and identifying low-lying bound states of proximitized hybrid devices \cite{Mourik2012,Deng2012,Das2012,Churchill2013,Finck2013,Deng2016,Chen2017,Nichele2017,Gul2018}, the regime above the superconducting gap has received less attention. In this regime we observe oscillations in the tunnel conductance which we believe arise from the influence of the 1D sub-bands of the etched quasi-1D InAs wire itself. Fitting a numerical model \cite{Zhang2009} supports this hypothesis and reveals the strength of the SOI in the confined 1D region -- a key parameter in evaluating the suitability of these systems for supporting Majorana bound states.

\begin{figure}[t]
\includegraphics[width=\linewidth]{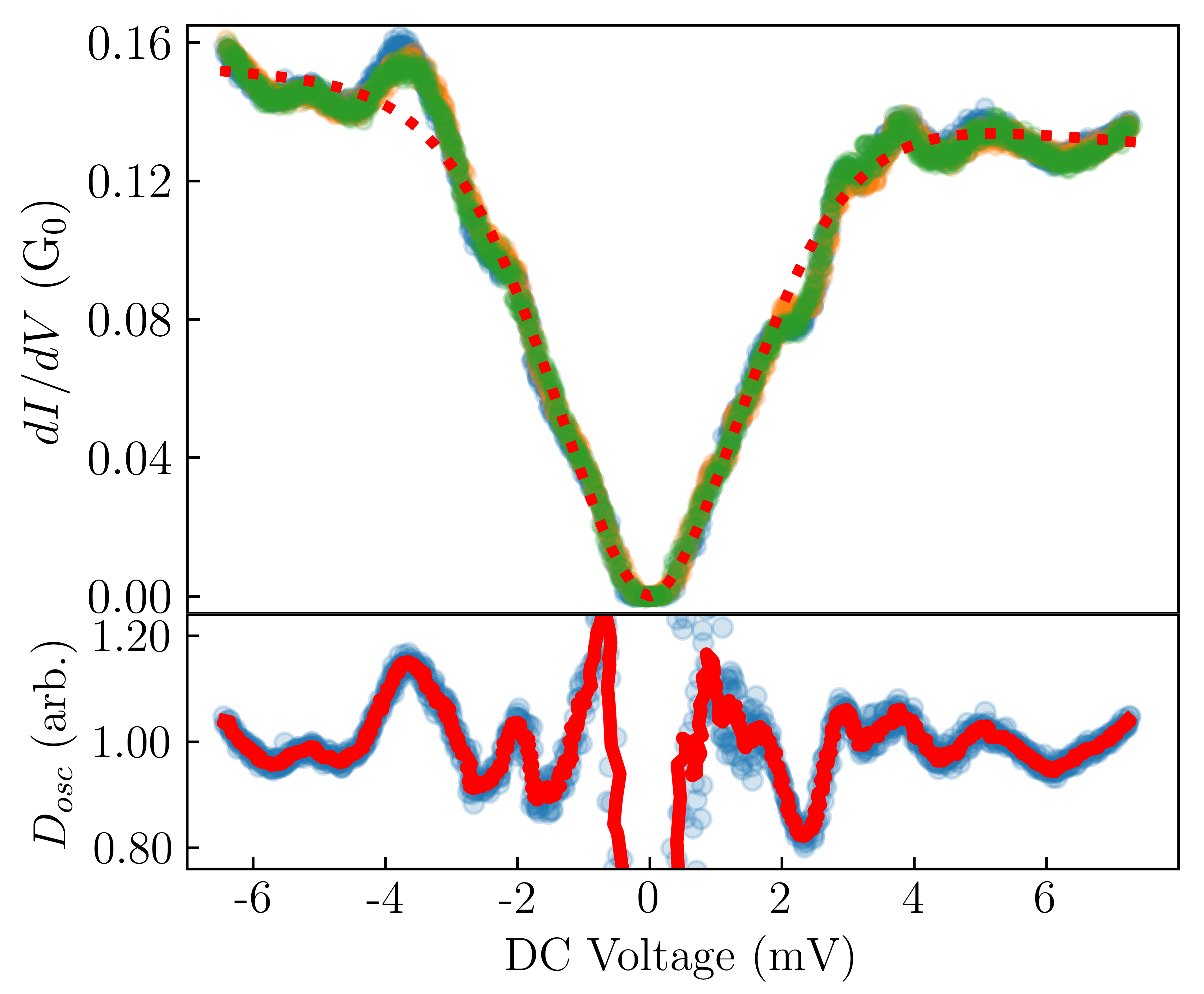}
\caption{\label{fig2} Upper panel: the measured differential conductance, $dI/dV$, as a function of bias at zero magnetic field (after x- [green], y- [yellow] and z-axis [blue] magnetic field sweeps). The red dotted line is a fit to the data described in the text. Lower panel: The blue circles are the oscillating part of the `wire' DOS, $D_{osc}$, extracted from the differential conductance as described in the text (red line) smoothed data.}
\end{figure}

\begin{figure*}
\includegraphics[width=\textwidth]{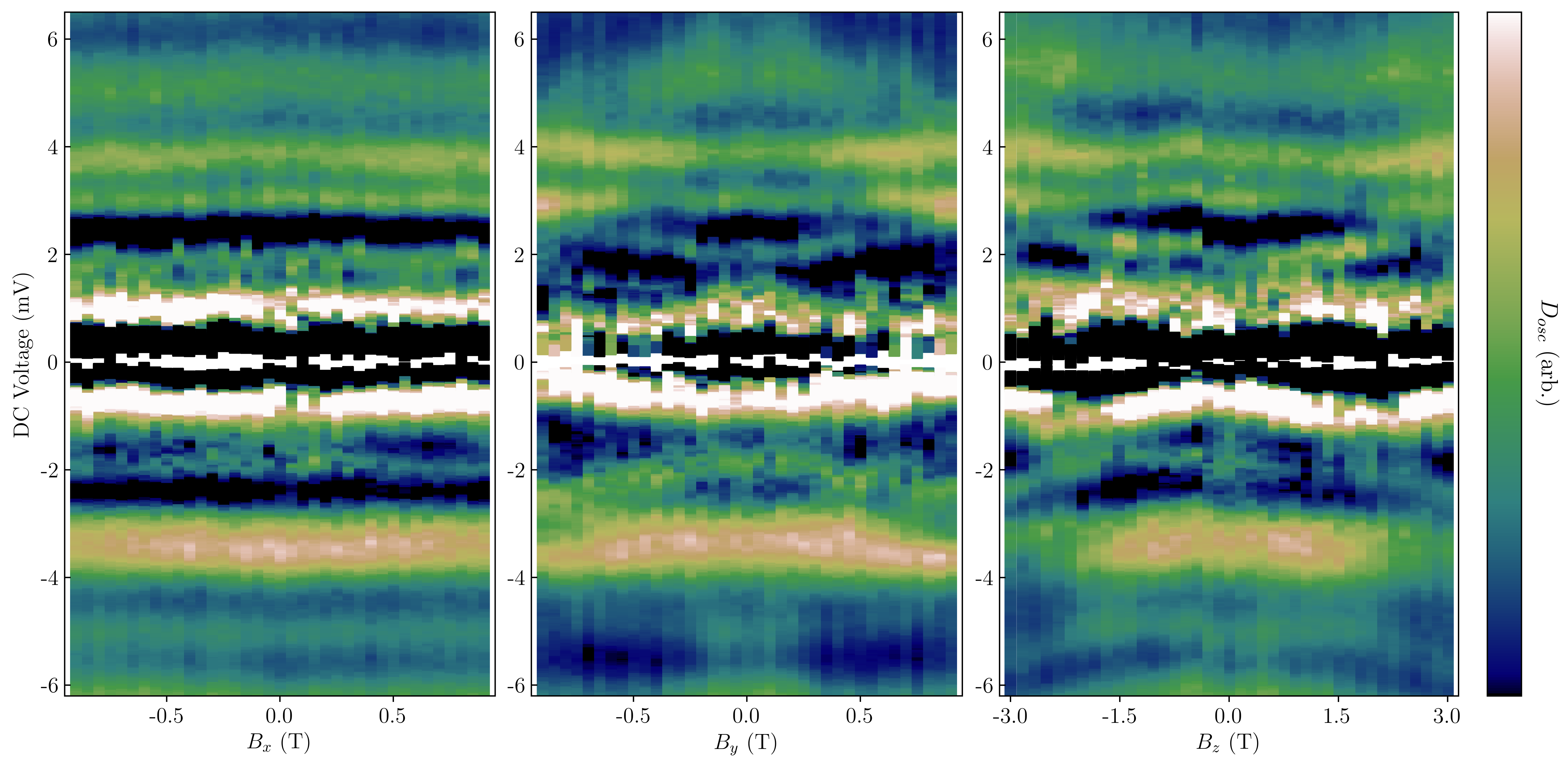}
\caption{\label{fig3} The magnetic field dependence of the fluctuating part of the measured DOS, $D_{osc}$, as a function of bias, shown for the three different magnetic field axes. How the axes relate to the device is shown in fig.~\ref{fig1}~(b).}
\end{figure*}

\section{Methods}

The InAs-Al hybrid heterostructure is grown by molecular beam epitaxy using ultra-high purity techniques \cite{Gardner2016}. The structure includes a complex quantum well of InAs and In$_{0.90}$Ga$_{0.10}$As grown on a lower barrier of In$_{0.81}$Al$_{0.19}$As. A 7~nm Al layer is deposited in-situ after a 2ML GaAs etch stop is deposited on the semiconductor structure. The tunnelling device, along with conventional Hall bar devices, were fabricated using a conventional e-beam lithography process, a phosphoric acid etch to define the mesa and a Transene etch to remove the aluminium layer from the 1D wire part of the device and from the active area of the Hall bars. Patterned and etched samples were then cleaned using the pulse--purge function of the ALD tool prior to depositing a dielectric layer and metallic top-gates \cite{Pauka2020}. 

A micrograph and schematic of the device are shown in fig.~\ref{fig1}~(a) and (b) respectively. It comprises, on the left hand side, a wide 2DEG mesa with a central, thin patterned Al strip on top, and on the right, a mesa-etched quasi-1D nanowire without Al on top. There is some variation of the wire width along its length, but near the tunnel barrier it is $280\pm90$~nm \footnote{It should also be noted that the measured visual width of the wire will exceed that of the 2DEG formed within -- from etch induced disorder and the confining potential}. On the left, three metallic top-gates can be seen. The outer top-gates serve to deplete the 2DEG in these regions only. The central top-gate -- which mirrors approximately the shape of the underlying patterned Al -- extends slightly (of order 20~nm) beyond the Al and thereby, when negatively biased, is able to deplete a thin region of the 2DEG (shown schematically in red). This forms a tunnel barrier between the Al side of the device and the quasi-1D nanowire side of the device. Transport measurements are performed using a standard 4-point lock-in technique with a DC offset bias. The sample is mounted in a dilution refrigerator equipped with a vector-magnet, operating at a temperature $< 10$~mK (electron temperature, $T_e \sim 55$~mK).

\section{Results}

The device is operated in the tunnelling regime -- the top-gate is used exclusively to form the tunnel barrier. At a bias of $-3$~V the conductance through the device is of order 0.2~$G_0$ or less. The side-gates were found to bear little influence on the experiment and were kept at 0~V. The Al superconducting gap is visible at low-bias before the application of the magnetic field and has a value of $\approx 250~\mu$eV [see supplementary information]. After the application of the magnetic fields the superconductivity did not recover and throughout this work we treat the Al as a normal 3D metal.


An example of the differential conductance, $dI/dV$, as a function of bias across the tunnel barrier can be seen in the upper panel of fig.~\ref{fig2}. The three different traces represent the zero magnetic field case taken after each axis field sweep -- fields were swept to their extreme positive value, to their extreme negative value and back to zero -- 1, 1 and 3~T for the x-, y- and z-axes, respectively. The similarity of the traces after each sweep shows the robustness of the band features after changing the bias and the magnetic field.

The current through a tunnel barrier is directly related to the density of electronic states, $D_{L,R}$, on each side by,

\begin{equation*}
    I \propto \int_{-\infty}^{\infty} T(E) D_L(E + eV) D_R(E) [f(E) - f(E + eV)] dE,
\end{equation*}

\noindent where $T$ is the tunneling coefficient and $f(E)$ is the Fermi-Dirac distribution at energy $E$. 

It is our hypothesis that the oscillatory features in the data correspond to the changing accessibility of the individual quasi-1D bands of the wire, that is, the peaks correspond to the (energetically broadened) van Hove singularities associated with the quasi-1D band minima \cite{Eugster1994}.

The measurement broadened quasi-1D DOS, $D_R$, can be decomposed into a background term, $D_{bg} \propto \sqrt{E-E_{bm}}$, where $E_{bm}$ is the effective ensemble band minimum, and an oscillating part, $D_{osc}$ (see SI). To isolate the features of interest in the data, $D_{osc}$, it is necessary to fit the remaining terms of the tunnelling integral, $D_{bg} D_L T(E)$ (a description and justification of the fit is given in the supplementary information)\cite{Devoret1990,SCT,Ukraintsev1996,Sonin2001,Tarkiainen2001,Waldecker2016}. This fit is shown in the upper panel of fig.~\ref{fig2} (red dotted line). 

The lower panel of fig.~\ref{fig2} shows $D_{osc}$ (blue circles) extracted from the data in the upper panel. The red curve is a smoothed version which is used throughout the subsequent magnetic field analysis. This method of extracting $D_{bg}$ renders the extracted $D_{osc}$ around zero-bias as undefined -- data points between $-300$ and $300~\mu$eV have therefore been excluded from the analysis. 

There is considerable, robust structure to the $D_{osc}$ data. The data display a somewhat periodic modulation with a peak spacing of $\sim 1$~meV. Although the periodicity of the peak positions in energy is somewhat regular, the magnitudes (occupancy) clearly differ.


Figure~\ref{fig3} shows the fluctuating part of the DOS, $D_{osc}$, as a function of applied magnetic field along three different axes. As above, the zero bias regime does not offer any meaningful information and has been excluded. The oscillations visible in fig.~\ref{fig2} are much more easily discerned when viewed as a function of magnetic field and there are distinct differences between the different axes. The sub-band structure appears largely insensitive to a magnetic field in the $x$-axis direction. Band deflection and band splitting, however, are visible for a magnetic field applied in both $y$- and $z$-axes (N.B.~due to the type of vector magnet used, the $z$-axis field range is greater than that of the $x$- and $y$-axes).

\section{Modelling}

To model the system, we followed the method of Zhang \emph{et al.}~\cite{Zhang2009} -- a numerical perturbative approach which includes the effects of the Rashba and Dresselhaus SOI as well as an SOI arising from the transverse confinement, along with the Zeeman effect. The model treats the transverse confinement (in $x$) as due to a parabolic potential and considers the application of a magnetic field along the $y$-axis only. 

\begin{figure}
\includegraphics[width=\linewidth]{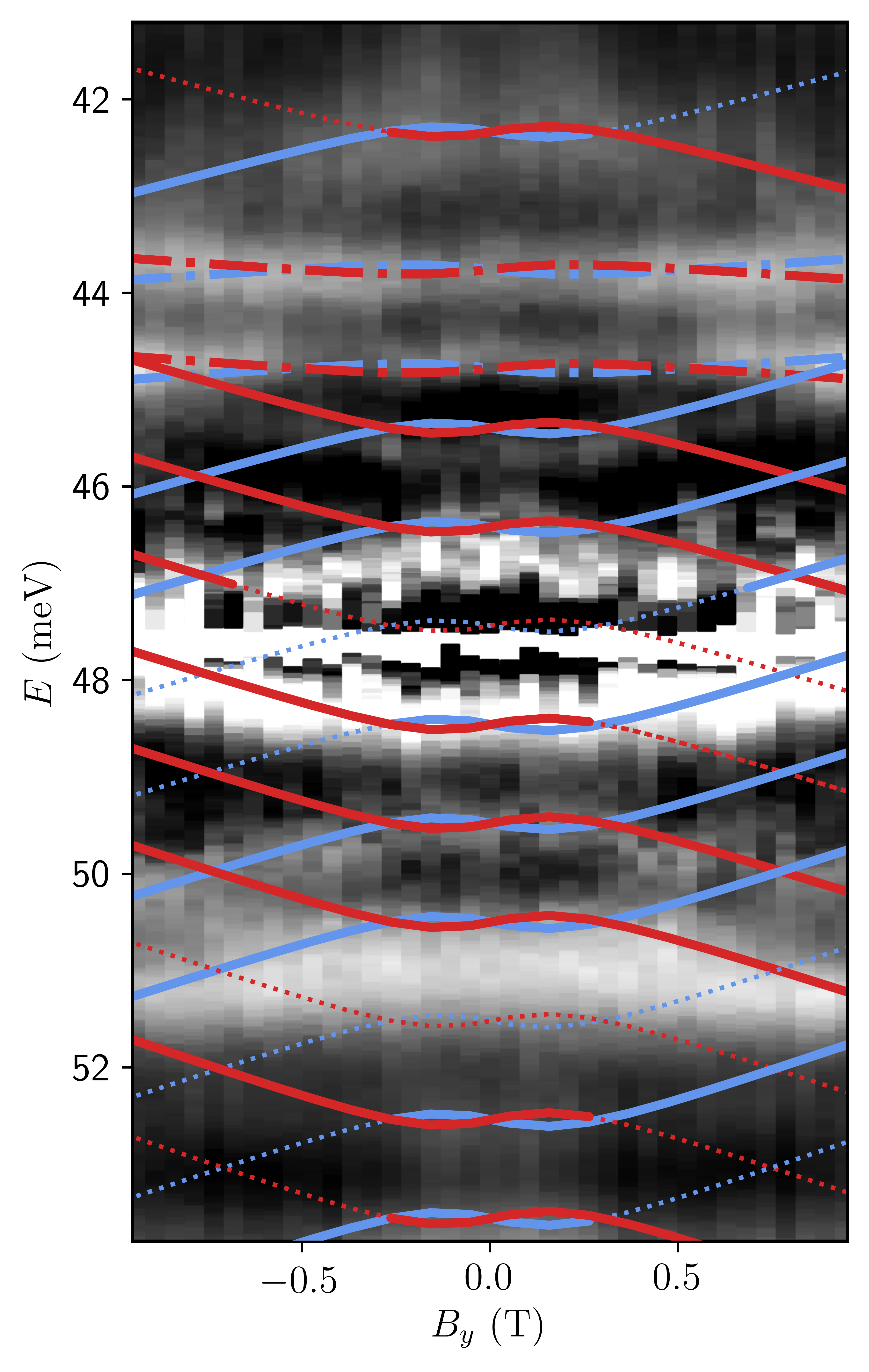}
\caption{\label{fig4} The magnetic field dependence of the fluctuating part of the measured DOS, $D_{osc}$, as a function of bias, for magnetic field in the $y$-axis direction. The energy is relative to the ensemble band minimum, $E_{bm}$. The red and blue lines are the different spin species of the modelled 1D sub-band structure. The solid lines are fits to the data, the dotted lines are band edges predicted by the model which have no obvious experimentally observed counterpart and the dot-dashed lines are fits which required modified fitting parameters.
}
\end{figure}

We calculate the dispersion relation for the different bands, $E(k)_n$, using this model, and then track the band minima as a function of applied magnetic field. These band minima along with the data from the central panel of fig.~\ref{fig3} (now in grey-scale) are shown in fig.~\ref{fig4}. The red and blue curves represent the different spin-bands. The solid and dotted lines are directly from the model -- the dotted lines show bands (or sections of bands) that do not have an obvious experimental equivalent in the data. The dot-dashed lines have different model parameters, discussed below and have been shifted in energy to better fit the data.

The fit shown was achieved with a Rashba parameter, $\alpha_R = 3.0 \times 10^{-12}$~eVm, the Dresselhaus parameter, $\beta_D = 0$~eVm, and the lateral spin-orbit parameter, $\gamma = 5.0 \times 10^{-13}$~eVm. The model was found to be largely insensitive to the g-factor in this field regime over the range of realistic values considered \cite{Lee2019}, as such, it was maintained at $g = 17$. With reference to the simple 1D DOS approximation (discussed in the SI) we considered the band number, $n$, to be centered about 48 (although due to the regular spacing of the bands, this choice is not critical to the analysis). The lateral confining potential is characterised by an angular frequency which we set as $\omega_0  = 1.55 \times 10^{12}$~Hz in order to achieve a level spacing comparable to that measured experimentally. This frequency corresponds to a length-scale, $l_0 \sim 50$~nm, which is of the order of the wire width, which is $280 \pm 90$~nm as measured (although the 2DEG itself will be confined to a narrower region due to disorder at the wire edge). To fit the dot-dashed curves, the Rashba term was reduced to $\alpha_R = 2.0 \times 10^{-12}$~eVm and the transverse SOI term to $\gamma = 2.0 \times 10^{-13}$~eVm, whilst everything else was kept as before.

\section{Discussion}

The model is able to reproduce the key features and dependencies present in the data with a very reasonable value for the $x$-axis parabolic confining potential. From measurements performed on comparable samples, $\alpha_R$ and $\beta_D$ are also reasonable and consistent for this type of stack \cite{Heida1998,Shojaei2016,Kaushini2018}. The lateral spin-orbit parameter, $\gamma$, is somewhat of an unknown quantity, but could reasonably be assumed to be of a similar order as the Rashba coupling parameter -- here we see it an order of magnitude smaller.

There are also, however, some irregularities across the measured range not accounted for by the model and it is apparent that several of the bands appear much more strongly in one spin or one directional splitting. There is also an apparently absent band (the third from the bottom) which seems possibly to be shifted into the adjacent band -- the brightest feature across this region. These inconsistencies suggest that there are possible additional terms that couple different bands which are not captured by this model. The reduction of the SOI and shifting in energy of only very specific and localised (in energy) bands at such a high band number, $n$, strongly suggests something akin to donor levels arising from impurities and coupling to bands at specific energies. It is also a possibility that non-parabolicity could account for this, although the regular spacing away from these shifted levels makes this an unlikely explanation.

\section{Conclusion}

In conclusion, we have demonstrated the feasibility of fabricating, using a subtractive process, a device possessed of quasi-1D characteristics and have shown the magnetic field dependence thereof. We were able to fit the data in one of the magnetic field directions with energy and SOI values that are consistent with the geometry of the device and the materials employed. Having shown that the 1D modes in a confined 2DEG are evenly spaced in energy, in principle, this allows one to address or occupy specific states of interest more easily and reliably than the mode structure of grown nanowires, which can be complex and highly degenerate. With the addition of an electrostatic gate to this system -- to adjust the energy of the bands in the mesa-etched 1D wire -- it is likely possible to move the wire from an even number of occupied bands to an odd number -- an essential ingredient for MBS generation. In addition, this technique also allows for the extraction of the strength of the SOI coupling parameters of a 1D system.

\section{Acknowledgements}

\
We wish to thank S. Pauka, A. Jouan, and M. Cassidy for useful conversations and technical help. This research was supported by the Microsoft Corporation and the Australian Research Council Centre of Excellence for Engineered Quantum Systems (EQUS, CE170100009). The authors acknowledge the facilities as well as the scientific and technical assistance of the Research \& Prototype Foundry Core Research Facility at the University of Sydney, part of the Australian National Fabrication Facility.

\bibliography{biblio}

\begin{thebibliography}{38}
\providecommand{\natexlab}[1]{#1}
\providecommand{\url}[1]{\texttt{#1}}
\expandafter\ifx\csname urlstyle\endcsname\relax
  \providecommand{\doi}[1]{doi: #1}\else
  \providecommand{\doi}{doi: \begingroup \urlstyle{rm}\Url}\fi

\bibitem[Lutchyn et~al.(2018)Lutchyn, Bakkers, Kouwenhoven, Krogstrup, Marcus,
  and Oreg]{Lutchyn2018}
R.~Lutchyn, E.~P. A.~M. Bakkers, L.~P. Kouwenhoven, P.~Krogstrup, C.~M. Marcus,
  and Y.~Oreg.
\newblock Majorana zero modes in superconductor–semiconductor
  heterostructure.
\newblock \emph{Nat. Rev. Mater.}, 3:\penalty0 52, 2018.
\newblock \doi{https://doi.org/10.1038/s41578-018-0003-1}.

\bibitem[Frolov et~al.(2020)Frolov, Manfra, and Sau]{Frolov2020}
S.~M. Frolov, M.~J. Manfra, and J.~D. Sau.
\newblock Topological superconductivity in hybrid devices.
\newblock \emph{Nature Physics}, 16\penalty0 (7):\penalty0 718--724, Jul 2020.
\newblock \doi{10.1038/s41567-020-0925-6}.

\bibitem[Mourik et~al.(2012)Mourik, Zuo, Frolov, Plissard, Bakkers, and
  Kouwenhoven]{Mourik2012}
V.~Mourik, K.~Zuo, S.~M. Frolov, S.~R. Plissard, E.~P. A.~M. Bakkers, and L.~P.
  Kouwenhoven.
\newblock Signatures of majorana fermions in hybrid
  superconductor-semiconductor nanowire devices.
\newblock \emph{Science}, 336\penalty0 (6084):\penalty0 1003--1007, 2012.
\newblock \doi{10.1126/science.1222360}.

\bibitem[Deng et~al.(2012)Deng, Yu, Huang, Larsson, Caroff, and Xu]{Deng2012}
M.~T. Deng, C.~L. Yu, G.~Y. Huang, M.~Larsson, P.~Caroff, and H.~Q. Xu.
\newblock Anomalous zero-bias conductance peak in a nb–insb nanowire–nb
  hybrid device.
\newblock \emph{Nano Letters}, 12\penalty0 (12):\penalty0 6414--6419, 2012.
\newblock \doi{10.1021/nl303758w}.

\bibitem[Das et~al.(2012)Das, Ronen, Most, Oreg, Heiblum, and
  Shtrikman]{Das2012}
Anindya Das, Yuval Ronen, Yonatan Most, Yuval Oreg, Moty Heiblum, and Hadas
  Shtrikman.
\newblock Zero-bias peaks and splitting in an al--inas nanowire topological
  superconductor as a signature of majorana fermions.
\newblock \emph{Nature Physics}, 8\penalty0 (12):\penalty0 887--895, Dec 2012.
\newblock \doi{10.1038/nphys2479}.

\bibitem[Churchill et~al.(2013)Churchill, Fatemi, Grove-Rasmussen, Deng,
  Caroff, Xu, and Marcus]{Churchill2013}
H.~O.~H. Churchill, V.~Fatemi, K.~Grove-Rasmussen, M.~T. Deng, P.~Caroff, H.~Q.
  Xu, and C.~M. Marcus.
\newblock Superconductor-nanowire devices from tunneling to the multichannel
  regime: Zero-bias oscillations and magnetoconductance crossover.
\newblock \emph{Phys. Rev. B}, 87:\penalty0 241401, Jun 2013.
\newblock \doi{10.1103/PhysRevB.87.241401}.

\bibitem[Finck et~al.(2013)Finck, Van~Harlingen, Mohseni, Jung, and
  Li]{Finck2013}
A.~D.~K. Finck, D.~J. Van~Harlingen, P.~K. Mohseni, K.~Jung, and X.~Li.
\newblock Anomalous modulation of a zero-bias peak in a hybrid
  nanowire-superconductor device.
\newblock \emph{Phys. Rev. Lett.}, 110:\penalty0 126406, Mar 2013.
\newblock \doi{10.1103/PhysRevLett.110.126406}.
\newblock URL \url{https://link.aps.org/doi/10.1103/PhysRevLett.110.126406}.

\bibitem[Deng et~al.(2016)Deng, Vaitiekenas, Hansen, Danon, Leijnse, Flensberg,
  Nyg{\r a}rd, Krogstrup, and Marcus]{Deng2016}
M.~T. Deng, S.~Vaitiekenas, E.~B. Hansen, J.~Danon, M.~Leijnse, K.~Flensberg,
  J.~Nyg{\r a}rd, P.~Krogstrup, and C.~M. Marcus.
\newblock Majorana bound state in a coupled quantum-dot hybrid-nanowire system.
\newblock \emph{Science}, 354\penalty0 (6319):\penalty0 1557--1562, 2016.
\newblock \doi{10.1126/science.aaf3961}.

\bibitem[Chen et~al.(2017)Chen, Yu, Stenger, Hocevar, Car, Plissard, Bakkers,
  Stanescu, and Frolov]{Chen2017}
Jun Chen, Peng Yu, John Stenger, Mo{\"\i}ra Hocevar, Diana Car,
  S{\'e}bastien~R. Plissard, Erik P. A.~M. Bakkers, Tudor~D. Stanescu, and
  Sergey~M. Frolov.
\newblock Experimental phase diagram of zero-bias conductance peaks in
  superconductor/semiconductor nanowire devices.
\newblock \emph{Science Advances}, 3\penalty0 (9), 2017.
\newblock \doi{10.1126/sciadv.1701476}.

\bibitem[G{\"u}l et~al.(2018)G{\"u}l, Zhang, Bommer, de~Moor, Car, Plissard,
  Bakkers, Geresdi, Watanabe, Taniguchi, and Kouwenhoven]{Gul2018}
{\"O}nder G{\"u}l, Hao Zhang, Jouri D.~S. Bommer, Michiel W.~A. de~Moor, Diana
  Car, S{\'e}bastien~R. Plissard, Erik P. A.~M. Bakkers, Attila Geresdi, Kenji
  Watanabe, Takashi Taniguchi, and Leo~P. Kouwenhoven.
\newblock Ballistic majorana nanowire devices.
\newblock \emph{Nature Nanotechnology}, 13\penalty0 (3):\penalty0 192--197, Mar
  2018.
\newblock \doi{10.1038/s41565-017-0032-8}.

\bibitem[Lee et~al.(2019{\natexlab{a}})Lee, Choi, Pendharkar, Pennachio,
  Markman, Seas, Koelling, Verheijen, Casparis, Petersson, Petkovic, Schaller,
  Rodwell, Marcus, Krogstrup, Kouwenhoven, Bakkers, and Palmstr\o{}m]{SAG1}
Joon~Sue Lee, Sukgeun Choi, Mihir Pendharkar, Daniel~J. Pennachio, Brian
  Markman, Michael Seas, Sebastian Koelling, Marcel~A. Verheijen, Lucas
  Casparis, Karl~D. Petersson, Ivana Petkovic, Vanessa Schaller, Mark J.~W.
  Rodwell, Charles~M. Marcus, Peter Krogstrup, Leo~P. Kouwenhoven, Erik P.
  A.~M. Bakkers, and Chris~J. Palmstr\o{}m.
\newblock Selective-area chemical beam epitaxy of in-plane inas one-dimensional
  channels grown on inp(001), inp(111)b, and inp(011) surfaces.
\newblock \emph{Phys. Rev. Materials}, 3:\penalty0 084606, Aug
  2019{\natexlab{a}}.
\newblock \doi{10.1103/PhysRevMaterials.3.084606}.
\newblock URL \url{https://link.aps.org/doi/10.1103/PhysRevMaterials.3.084606}.

\bibitem[Vaitiek\ifmmode~\dot{e}\else \.{e}\fi{}nas
  et~al.(2018)Vaitiek\ifmmode~\dot{e}\else \.{e}\fi{}nas, Whiticar, Deng,
  Krizek, Sestoft, Palmstr\o{}m, Marti-Sanchez, Arbiol, Krogstrup, Casparis,
  and Marcus]{SAG2}
S.~Vaitiek\ifmmode~\dot{e}\else \.{e}\fi{}nas, A.~M. Whiticar, M.-T. Deng,
  F.~Krizek, J.~E. Sestoft, C.~J. Palmstr\o{}m, S.~Marti-Sanchez, J.~Arbiol,
  P.~Krogstrup, L.~Casparis, and C.~M. Marcus.
\newblock Selective-area-grown semiconductor-superconductor hybrids: A basis
  for topological networks.
\newblock \emph{Phys. Rev. Lett.}, 121:\penalty0 147701, Oct 2018.
\newblock \doi{10.1103/PhysRevLett.121.147701}.
\newblock URL \url{https://link.aps.org/doi/10.1103/PhysRevLett.121.147701}.

\bibitem[Op~het Veld et~al.(2020)Op~het Veld, Xu, Schaller, Verheijen, Peters,
  Jung, Tong, Wang, de~Moor, Hesselmann, Vermeulen, Bommer, Sue~Lee, Sarikov,
  Pendharkar, Marzegalli, Koelling, Kouwenhoven, Miglio, Palmstr{\o}m, Zhang,
  and Bakkers]{SAG3}
Roy L.~M. Op~het Veld, Di~Xu, Vanessa Schaller, Marcel~A. Verheijen, Stan M.~E.
  Peters, Jason Jung, Chuyao Tong, Qingzhen Wang, Michiel W.~A. de~Moor, Bart
  Hesselmann, Kiefer Vermeulen, Jouri D.~S. Bommer, Joon Sue~Lee, Andrey
  Sarikov, Mihir Pendharkar, Anna Marzegalli, Sebastian Koelling, Leo~P.
  Kouwenhoven, Leo Miglio, Chris~J. Palmstr{\o}m, Hao Zhang, and Erik P. A.~M.
  Bakkers.
\newblock In-plane selective area insb--al nanowire quantum networks.
\newblock \emph{Communications Physics}, 3\penalty0 (1):\penalty0 59, Mar 2020.
\newblock ISSN 2399-3650.
\newblock \doi{10.1038/s42005-020-0324-4}.
\newblock URL \url{https://doi.org/10.1038/s42005-020-0324-4}.

\bibitem[Aseev et~al.(2019{\natexlab{a}})Aseev, Fursina, Boekhout, Krizek,
  Sestoft, Borsoi, Heedt, Wang, Binci, Mart{\'i}-S{\'a}nchez, Swoboda, Koops,
  Uccelli, Arbiol, Krogstrup, Kouwenhoven, and Caroff]{SAG4}
Pavel Aseev, Alexandra Fursina, Frenk Boekhout, Filip Krizek, Joachim~E.
  Sestoft, Francesco Borsoi, Sebastian Heedt, Guanzhong Wang, Luca Binci, Sara
  Mart{\'i}-S{\'a}nchez, Timm Swoboda, Ren{\'e} Koops, Emanuele Uccelli, Jordi
  Arbiol, Peter Krogstrup, Leo~P. Kouwenhoven, and Philippe Caroff.
\newblock Selectivity map for molecular beam epitaxy of advanced iii--v quantum
  nanowire networks.
\newblock \emph{Nano Letters}, 19\penalty0 (1):\penalty0 218--227, Jan
  2019{\natexlab{a}}.
\newblock ISSN 1530-6984.
\newblock \doi{10.1021/acs.nanolett.8b03733}.
\newblock URL \url{https://doi.org/10.1021/acs.nanolett.8b03733}.

\bibitem[Aseev et~al.(2019{\natexlab{b}})Aseev, Wang, Binci, Singh,
  Mart{\'i}-S{\'a}nchez, Botifoll, Stek, Bordin, Watson, Boekhout, Abel,
  Gamble, Van~Hoogdalem, Arbiol, Kouwenhoven, de~Lange, and Caroff]{SAG5}
Pavel Aseev, Guanzhong Wang, Luca Binci, Amrita Singh, Sara
  Mart{\'i}-S{\'a}nchez, Marc Botifoll, Lieuwe~J. Stek, Alberto Bordin, John~D.
  Watson, Frenk Boekhout, Daniel Abel, John Gamble, Kevin Van~Hoogdalem, Jordi
  Arbiol, Leo~P. Kouwenhoven, Gijs de~Lange, and Philippe Caroff.
\newblock Ballistic insb nanowires and networks via metal-sown selective area
  growth.
\newblock \emph{Nano Letters}, 19\penalty0 (12):\penalty0 9102--9111, Dec
  2019{\natexlab{b}}.
\newblock ISSN 1530-6984.
\newblock \doi{10.1021/acs.nanolett.9b04265}.
\newblock URL \url{https://doi.org/10.1021/acs.nanolett.9b04265}.

\bibitem[Desplanque et~al.(2018)Desplanque, Bucamp, Troadec, Patriarche, and
  Wallart]{SAG6}
L~Desplanque, A~Bucamp, D~Troadec, G~Patriarche, and X~Wallart.
\newblock In-plane {InSb} nanowires grown by selective area molecular beam
  epitaxy on semi-insulating substrate.
\newblock \emph{Nanotechnology}, 29\penalty0 (30):\penalty0 305705, may 2018.
\newblock \doi{10.1088/1361-6528/aac321}.
\newblock URL \url{https://doi.org/10.1088/1361-6528/aac321}.

\bibitem[Kjaergaard et~al.(2016)Kjaergaard, Nichele, Suominen, Nowak, Wimmer,
  Akhmerov, Folk, Flensberg, Shabani, Palmstr{\o}m, and Marcus]{Kjaergaard2016}
M.~Kjaergaard, F.~Nichele, H.~J. Suominen, M.~P. Nowak, M.~Wimmer, A.~R.
  Akhmerov, J.~A. Folk, K.~Flensberg, J.~Shabani, C.~J. Palmstr{\o}m, and C.~M.
  Marcus.
\newblock Quantized conductance doubling and hard gap in a two-dimensional
  semiconductor--superconductor heterostructure.
\newblock \emph{Nature Communications}, 7\penalty0 (1):\penalty0 12841, Sep
  2016.
\newblock ISSN 2041-1723.
\newblock \doi{10.1038/ncomms12841}.
\newblock URL \url{https://doi.org/10.1038/ncomms12841}.

\bibitem[Shabani et~al.(2016)Shabani, Kjaergaard, Suominen, Kim, Nichele,
  Pakrouski, Stankevic, Lutchyn, Krogstrup, Feidenhans'l, Kraemer, Nayak,
  Troyer, Marcus, and Palmstr\o{}m]{Shabani2016}
J.~Shabani, M.~Kjaergaard, H.~J. Suominen, Younghyun Kim, F.~Nichele,
  K.~Pakrouski, T.~Stankevic, R.~M. Lutchyn, P.~Krogstrup, R.~Feidenhans'l,
  S.~Kraemer, C.~Nayak, M.~Troyer, C.~M. Marcus, and C.~J. Palmstr\o{}m.
\newblock Two-dimensional epitaxial superconductor-semiconductor
  heterostructures: A platform for topological superconducting networks.
\newblock \emph{Phys. Rev. B}, 93:\penalty0 155402, Apr 2016.
\newblock \doi{10.1103/PhysRevB.93.155402}.
\newblock URL \url{https://link.aps.org/doi/10.1103/PhysRevB.93.155402}.

\bibitem[Nichele et~al.(2017)Nichele, Drachmann, Whiticar, O'Farrell, Suominen,
  Fornieri, Wang, Gardner, Thomas, Hatke, Krogstrup, Manfra, Flensberg, and
  Marcus]{Nichele2017}
Fabrizio Nichele, Asbj\o{}rn C.~C. Drachmann, Alexander~M. Whiticar, Eoin C.~T.
  O'Farrell, Henri~J. Suominen, Antonio Fornieri, Tian Wang, Geoffrey~C.
  Gardner, Candice Thomas, Anthony~T. Hatke, Peter Krogstrup, Michael~J.
  Manfra, Karsten Flensberg, and Charles~M. Marcus.
\newblock Scaling of majorana zero-bias conductance peaks.
\newblock \emph{Phys. Rev. Lett.}, 119:\penalty0 136803, Sep 2017.
\newblock \doi{10.1103/PhysRevLett.119.136803}.

\bibitem[Suominen et~al.(2017)Suominen, Kjaergaard, Hamilton, Shabani,
  Palmstr\o{}m, Marcus, and Nichele]{Suominen2017}
H.~J. Suominen, M.~Kjaergaard, A.~R. Hamilton, J.~Shabani, C.~J. Palmstr\o{}m,
  C.~M. Marcus, and F.~Nichele.
\newblock Zero-energy modes from coalescing andreev states in a two-dimensional
  semiconductor-superconductor hybrid platform.
\newblock \emph{Phys. Rev. Lett.}, 119:\penalty0 176805, Oct 2017.
\newblock \doi{10.1103/PhysRevLett.119.176805}.
\newblock URL \url{https://link.aps.org/doi/10.1103/PhysRevLett.119.176805}.

\bibitem[Plissard et~al.(2012)Plissard, Slapak, Verheijen, Hocevar, Immink, van
  Weperen, Nadj-Perge, Frolov, Kouwenhoven, and Bakkers]{VLS1}
S{\'e}bastien~R. Plissard, Dorris~R. Slapak, Marcel~A. Verheijen, Mo{\"i}ra
  Hocevar, George W.~G. Immink, Ilse van Weperen, Stevan Nadj-Perge, Sergey~M.
  Frolov, Leo~P. Kouwenhoven, and Erik P. A.~M. Bakkers.
\newblock From insb nanowires to nanocubes: Looking for the sweet spot.
\newblock \emph{Nano Letters}, 12\penalty0 (4):\penalty0 1794--1798, Apr 2012.
\newblock ISSN 1530-6984.
\newblock \doi{10.1021/nl203846g}.
\newblock URL \url{https://doi.org/10.1021/nl203846g}.

\bibitem[Car et~al.(2014)Car, Wang, Verheijen, Bakkers, and Plissard]{VLS2}
Diana Car, Jia Wang, Marcel~A. Verheijen, Erik P. A.~M. Bakkers, and
  Sébastien~R. Plissard.
\newblock Rationally designed single-crystalline nanowire networks.
\newblock \emph{Advanced Materials}, 26\penalty0 (28):\penalty0 4875--4879,
  2014.
\newblock \doi{https://doi.org/10.1002/adma.201400924}.
\newblock URL
  \url{https://onlinelibrary.wiley.com/doi/abs/10.1002/adma.201400924}.

\bibitem[Hatke et~al.(2017)Hatke, Wang, Thomas, Gardner, and Manfra]{Manfra}
A.~T. Hatke, T.~Wang, C.~Thomas, G.~C. Gardner, and M.~J. Manfra.
\newblock Mobility in excess of $10^6 \textrm{cm}^2/\textrm{V s}$ in inas
  quantum wells grown on lattice mismatched inp substrates.
\newblock \emph{Applied Physics Letters}, 111\penalty0 (14):\penalty0 142106,
  2017.
\newblock \doi{10.1063/1.4993784}.

\bibitem[Pauka et~al.(2020)Pauka, Witt, Allen, Harlech-Jones, Jouan, Gardner,
  Gronin, Wang, Thomas, Manfra, Gukelberger, Gamble, Reilly, and
  Cassidy]{Pauka2020}
S.~J. Pauka, J.~D.~S. Witt, C.~N. Allen, B.~Harlech-Jones, A.~Jouan, G.~C.
  Gardner, S.~Gronin, T.~Wang, C.~Thomas, M.~J. Manfra, J.~Gukelberger,
  J.~Gamble, D.~J. Reilly, and M.~C. Cassidy.
\newblock Repairing the surface of inas-based topological heterostructures.
\newblock \emph{Journal of Applied Physics}, 128:\penalty0 114301, 2020.
\newblock \doi{https://doi.org/10.1063/5.0014361}.

\bibitem[Zhang et~al.(2009)Zhang, Zhao, and Liu]{Zhang2009}
Tong-Yi Zhang, Wei Zhao, and Xue-Ming Liu.
\newblock Energy dispersion of the electrosubbands in parabolic confining
  quantum wires: interplay of rashba, dresselhaus, lateral spin–orbit
  interaction and the zeeman effect.
\newblock \emph{J. Phys.: Condens. Matter}, 21:\penalty0 335501, 2009.
\newblock \doi{10.1088/0953-8984/21/33/335501}.

\bibitem[Gardner et~al.(2016)Gardner, Fallahi, Watson, and Manfra]{Gardner2016}
G.~C. Gardner, S.~Fallahi, J.~D. Watson, and M.~J. Manfra.
\newblock Modified mbe hardware and techniques and role of gallium purity for
  attainment of two dimensional electron gas mobility $> 35 \times 10^6
  \textrm{cm}^2/\textrm{Vs}$ in algaas/gaas quantum wells grown by mbe.
\newblock \emph{Journal of Crystal Growth}, 441:\penalty0 71 -- 77, 2016.
\newblock ISSN 0022-0248.
\newblock \doi{https://doi.org/10.1016/j.jcrysgro.2016.02.010}.

\bibitem[Note1()]{Note1}
Note1.
\newblock It should also be noted that the measured visual width of the wire
  will exceed that of the 2DEG formed within -- from etch induced disorder and
  the confining potential.

\bibitem[Eugster et~al.(1994)Eugster, del Alamo, Rooks, and
  Melloch]{Eugster1994}
C.~C. Eugster, J.~A. del Alamo, M.~J. Rooks, and M.~R. Melloch.
\newblock One‐dimensional to one‐dimensional tunnelling between electron
  waveguides.
\newblock \emph{Applied Physics Letters}, 64\penalty0 (23):\penalty0
  3157--3159, 1994.
\newblock \doi{10.1063/1.111324}.
\newblock URL \url{https://doi.org/10.1063/1.111324}.

\bibitem[Devoret et~al.(1990)Devoret, Esteve, Grabert, Ingold, Pothier, and
  Urbina]{Devoret1990}
M.~H. Devoret, D.~Esteve, H.~Grabert, G.-L. Ingold, H.~Pothier, and C.~Urbina.
\newblock Effect of the electromagnetic environment on the coulomb blockade in
  ultrasmall tunnel junctions.
\newblock \emph{Phys. Rev. Lett.}, 64:\penalty0 1824--1827, Apr 1990.
\newblock \doi{10.1103/PhysRevLett.64.1824}.
\newblock URL \url{https://link.aps.org/doi/10.1103/PhysRevLett.64.1824}.

\bibitem[Grabert and Devoret(1992)]{SCT}
Hermann Grabert and Michel~H. Devoret, editors.
\newblock \emph{Single Charge Tunneling: Coulomb Blockade Phenomena In
  Nanostructures}.
\newblock Springer, Boston, MA, 1992.
\newblock \doi{https://doi.org/10.1007/978-1-4757-2166-9}.

\bibitem[Ukraintsev(1996)]{Ukraintsev1996}
Vladimir~A. Ukraintsev.
\newblock Data evaluation technique for electron-tunneling spectroscopy.
\newblock \emph{Phys. Rev. B}, 53:\penalty0 11176--11185, Apr 1996.
\newblock \doi{10.1103/PhysRevB.53.11176}.
\newblock URL \url{https://link.aps.org/doi/10.1103/PhysRevB.53.11176}.

\bibitem[Sonin(2001)]{Sonin2001}
E.~B. Sonin.
\newblock Tunneling into 1d and quasi-1d conductors and luttinger-liquid
  behavior.
\newblock \emph{Journal of Low Temperature Physics}, 124\penalty0 (1):\penalty0
  321--334, Jul 2001.
\newblock ISSN 1573-7357.
\newblock \doi{10.1023/A:1017598423240}.
\newblock URL \url{https://doi.org/10.1023/A:1017598423240}.

\bibitem[Tarkiainen et~al.(2001)Tarkiainen, Ahlskog, Penttil\"a, Roschier,
  Hakonen, Paalanen, and Sonin]{Tarkiainen2001}
R.~Tarkiainen, M.~Ahlskog, J.~Penttil\"a, L.~Roschier, P.~Hakonen, M.~Paalanen,
  and E.~Sonin.
\newblock Multiwalled carbon nanotube: Luttinger versus fermi liquid.
\newblock \emph{Phys. Rev. B}, 64:\penalty0 195412, Oct 2001.
\newblock \doi{10.1103/PhysRevB.64.195412}.
\newblock URL \url{https://link.aps.org/doi/10.1103/PhysRevB.64.195412}.

\bibitem[Waldecker et~al.(2016)Waldecker, Bertoni, Ernstorfer, and
  Vorberger]{Waldecker2016}
Lutz Waldecker, Roman Bertoni, Ralph Ernstorfer, and Jan Vorberger.
\newblock Electron-phonon coupling and energy flow in a simple metal beyond the
  two-temperature approximation.
\newblock \emph{Phys. Rev. X}, 6:\penalty0 021003, Apr 2016.
\newblock \doi{10.1103/PhysRevX.6.021003}.
\newblock URL \url{https://link.aps.org/doi/10.1103/PhysRevX.6.021003}.

\bibitem[Lee et~al.(2019{\natexlab{b}})Lee, Shojaei, Pendharkar, McFadden, Kim,
  Suominen, Kjaergaard, Nichele, Zhang, Marcus, and Palmstrøm]{Lee2019}
Joon~Sue Lee, Borzoyeh Shojaei, Mihir Pendharkar, Anthony~P. McFadden,
  Younghyun Kim, Henri~J. Suominen, Morten Kjaergaard, Fabrizio Nichele, Hao
  Zhang, Charles~M. Marcus, and Chris~J. Palmstrøm.
\newblock Transport studies of epi-al/inas two-dimensional electron gas systems
  for required building-blocks in topological superconductor networks.
\newblock \emph{Nano Letters}, 19\penalty0 (5):\penalty0 3083--3090,
  2019{\natexlab{b}}.
\newblock \doi{10.1021/acs.nanolett.9b00494}.

\bibitem[Heida et~al.(1998)Heida, van Wees, Kuipers, Klapwijk, and
  Borghs]{Heida1998}
J.~P. Heida, B.~J. van Wees, J.~J. Kuipers, T.~M. Klapwijk, and G.~Borghs.
\newblock Spin-orbit interaction in a two-dimensional electron gas in a
  inas/alsb quantum well with gate-controlled electron density.
\newblock \emph{Phys. Rev. B}, 57:\penalty0 11911--11914, May 1998.
\newblock \doi{10.1103/PhysRevB.57.11911}.
\newblock URL \url{https://link.aps.org/doi/10.1103/PhysRevB.57.11911}.

\bibitem[Shojaei et~al.(2016)Shojaei, O'Malley, Shabani, Roushan, Schultz,
  Lutchyn, Nayak, Martinis, and Palmstr{\o}m]{Shojaei2016}
B.~Shojaei, P.~J.~J. O'Malley, J.~Shabani, P.~Roushan, B.~D. Schultz, R.~M.
  Lutchyn, C.~Nayak, J.~M. Martinis, and C.~J. Palmstr{\o}m.
\newblock Demonstration of gate control of spin splitting in a high-mobility
  inas/alsb two-dimensional electron gas.
\newblock \emph{Phys. Rev. B}, 93:\penalty0 075302, Feb 2016.
\newblock \doi{10.1103/PhysRevB.93.075302}.
\newblock URL \url{https://link.aps.org/doi/10.1103/PhysRevB.93.075302}.

\bibitem[Wickramasinghe et~al.(2018)Wickramasinghe, Mayer, Yuan, Nguyen, Jiao,
  Manucharyan, and Shabani]{Kaushini2018}
Kaushini~S. Wickramasinghe, William Mayer, Joseph Yuan, Tri Nguyen, Lucy Jiao,
  Vladimir Manucharyan, and Javad Shabani.
\newblock Transport properties of near surface inas two-dimensional
  heterostructures.
\newblock \emph{Appl. Phys. Lett.}, 113:\penalty0 262104, 2018.

\end{thebibliography}

\end{document}


\title{Supplementary Information: Spin-orbit Energies in Etch-Confined Superconductor-Semiconductor Nanowires}

\author{J.~D.~S.~Witt}
\email{james.witt@sydney.edu.au}
\affiliation{ARC Centre of Excellence for Engineered Quantum Systems, School of Physics, The University of Sydney, Sydney, NSW 2006, Australia}

\author{G.~C.~Gardner}
\affiliation{Department of Physics and Astronomy, Purdue University, West Lafayette, Indiana, USA}
\affiliation{Microsoft Quantum Purdue, Purdue University, West Lafayette, Indiana, USA}

\author{C.~Thomas} 
\affiliation{Department of Physics and Astronomy, Purdue University, West Lafayette, Indiana, USA}
\affiliation{Microsoft Quantum Purdue, Purdue University, West Lafayette, Indiana, USA}

\author{T.~Lindemann} 
\affiliation{Department of Physics and Astronomy, Purdue University, West Lafayette, Indiana, USA}
\affiliation{Microsoft Quantum Purdue, Purdue University, West Lafayette, Indiana, USA}

\author{S.~Gronin} 
\affiliation{Department of Physics and Astronomy, Purdue University, West Lafayette, Indiana, USA}
\affiliation{Microsoft Quantum Purdue, Purdue University, West Lafayette, Indiana, USA}

\author{M.~J.~Manfra}
\affiliation{Department of Physics and Astronomy, Purdue University, West Lafayette, Indiana, USA}
\affiliation{Microsoft Quantum Purdue, Purdue University, West Lafayette, Indiana, USA}

\author{D.~J.~Reilly}
\email{david.reilly@sydney.edu.au}
\affiliation{ARC Centre of Excellence for Engineered Quantum Systems, School of Physics, The University of Sydney, Sydney, NSW 2006, Australia}
\affiliation{Microsoft Quantum Sydney, The University of Sydney, Sydney, NSW 2006, Australia}
\date{\today}

\maketitle

\section{Hall Measurements}

From the (top-gated) Hall bar, patterned on the same chip, the electron density was found to be between $0.5-2.5 \times 10^{12}~\textrm{cm}^{-2}$ with a mobility of between $3,000-7,000~\textrm{cm}^{2}\textrm{V}^{-1}\textrm{s}^{-1}$.



\section{Superconducting Gap}

\begin{figure}[h]
\includegraphics[width=10cm]{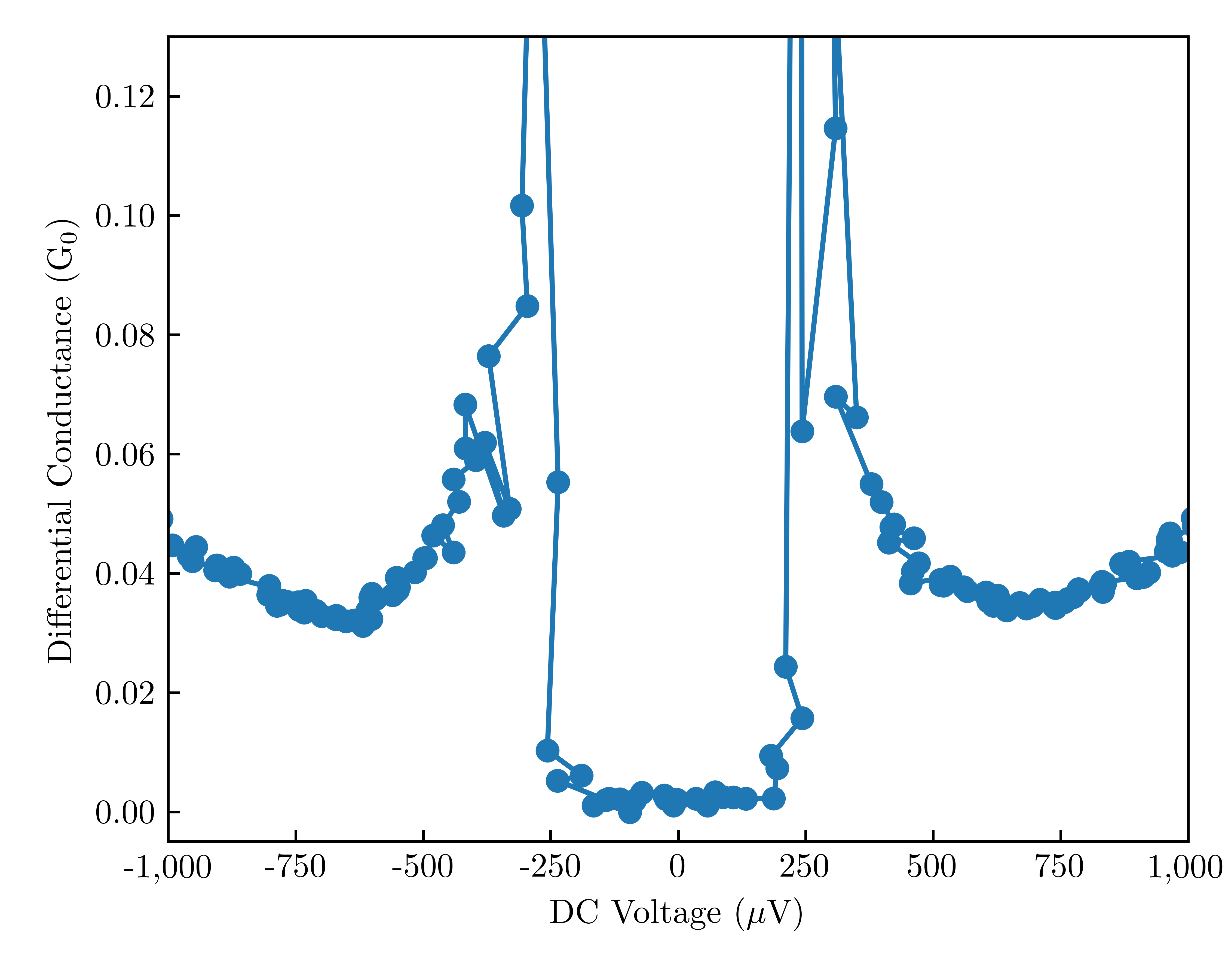}
\caption{\label{figS1} }
\end{figure}

Figure~\ref{figS1} shows the superconducting gap in the low-bias regime. The values of the gap, $\Delta \sim 250~\mu$eV.

\section{1D DOS Approximation}

\begin{figure}[h]
\includegraphics[width=\textwidth]{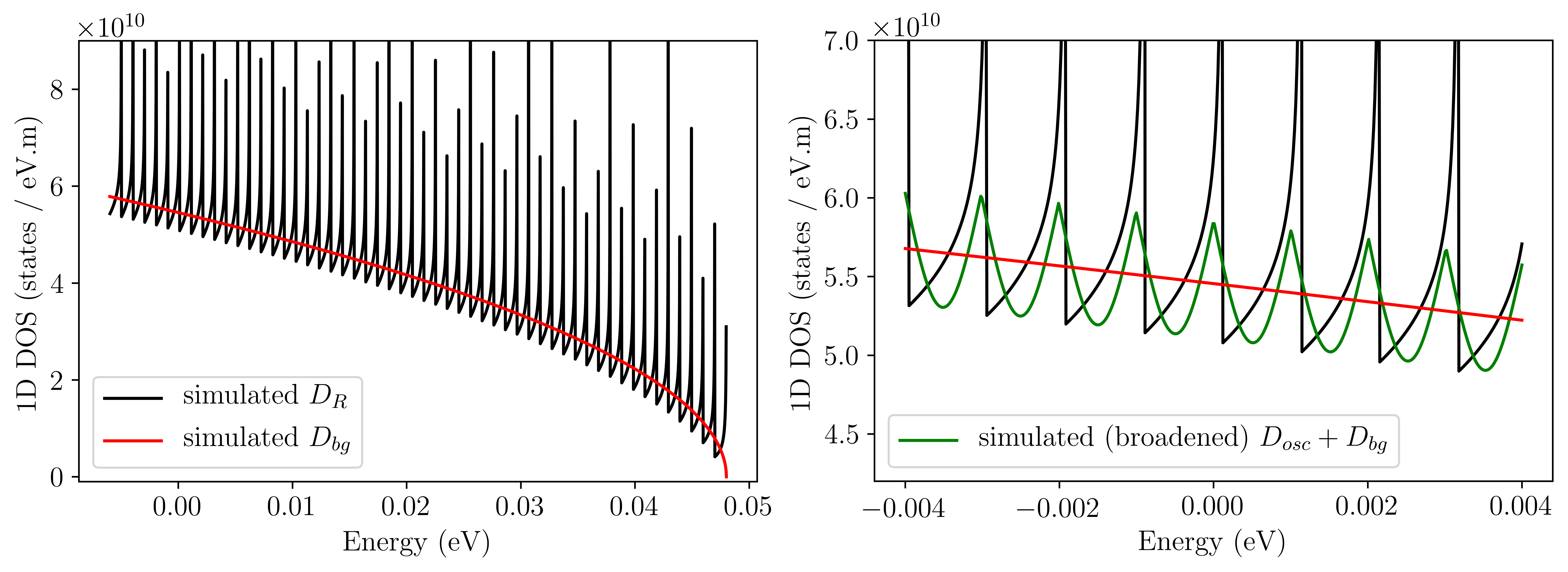}
\caption{\label{figS2} }
\end{figure}

Figure~\ref{figS2}~(left) shows the simulated ideal quasi-1D DOS (black line) for an idealised parabolic potential, equivalent to equally spaced bands with individual 1D DOS $\propto 1/\sqrt{E}$. The overall trend can be well approximated by, $D_{bg} \propto \sqrt{E-E_{bm}}$ (red line). The irregularity in the height of the Van Hove singularities is an artefact of the plot resolution. The band ensemble minimum, $E_{bm} = 48$~meV, for a discussion of this value see section~\ref{here}.

The thermal broadening in our system is only $\approx 5~\mu$eV, the smoothing of the features that we observe most likely arises from a combination of this and the lock-in excitation $\approx 100~\mu$eV.

Figure~\ref{figS2}~(right) shows an expanded section of the left panel around the Fermi energy. A simulated $D_{bg} + D_{osc}$, where $D_{osc} \propto \lvert \sin (\lambda_V) \rvert$, is shown as an example of broadening (green line).

\section{2D $\rightarrow$ Quasi-1D}\label{here}

The band minimum in the quasi-1D DOS approximation is set so that the total electron number calculated from the occupation of the 1D bands up to the Fermi energy is consistent with the measured 2D electron density when approaching the quasi-1D limit. Considering the errors in wire width and measured 2D electron density, the band minimum, $E_{bm} = 48\pm13$~meV. 

\section{Fitting $D_{bg} D_L T(E)$}

\begin{figure}[h]
\includegraphics[width=\textwidth]{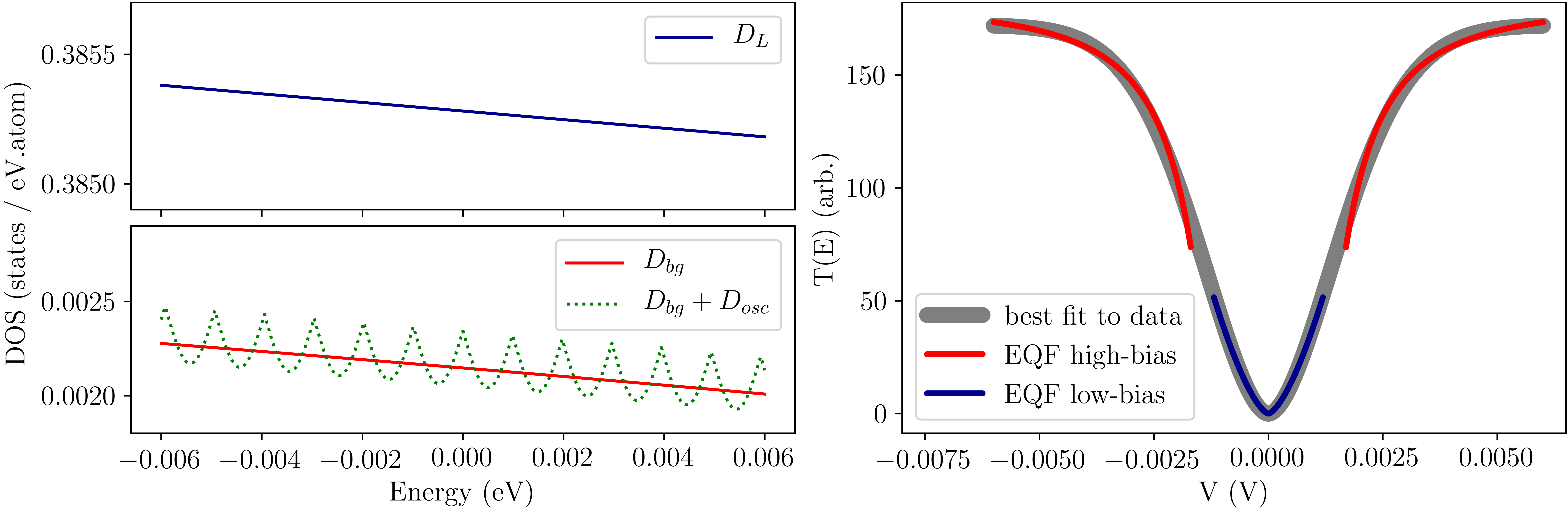}
\caption{\label{figS3} }
\end{figure}

To isolate the oscillatory features in the data which arise from the peaked DOS at the quasi-1D band minima, $D_{osc}$, it is neccessary to fit the other terms in the tunnelling relation. The form used for $D_{bg}$ is given above and is shown in fig.~\ref{figS3} (with a change of units) along with $D_{L}$ -- considered as arising from an ideal 3D metal with a Fermi energy of $11.6$~eV -- for Al this is a reasonable assumption~\cite{Waldecker2016}. It should be noted that the percentage change of states over the applicable energy range (shown) is much more significant for the wire than for the 3D metal. Hence, although the wire DOS is much lower, it has an appreciable effect on the tunnelling conductance. 

The actual conductance will be scaled by the number of atomic sites participating in the tunnelling in conjunction with the transmission coefficient, $T(E)$, which also defines the characteristic tunneling conductance shape. 

To describe the transmission coefficient in tunnelling from a 3D metal to a quasi-1D system the Environment Quantum Fluctuation (EQF) theory should be used \cite{Devoret1990,SCT,Sonin2001,Tarkiainen2001}. At low-bias -- when the voltage, $V < e/C_{T}$, where $C_{T}$ is the tunnel junction capacitance -- the model describes the \emph{Coloumb blockade} regime. In this regime the differential conductance has a power-law dependence on voltage,

\begin{equation*}
    \frac{dI}{dV} \propto V^{\alpha_E}
\end{equation*}

\noindent where $\alpha_E = 2Z/R_{K}$, $Z$ is the impedance of the circuit environment, and $R_{K} = 25,813~\Omega$ is the von Klitzing constant. It should be noted that in this analysis all of the DOS terms are absorbed into the constant of proportionality, what remains is a combination of $T(E)$ and the distribution functions. These terms are grouped under the $T(E)$ term here. From fitting the data in this regime (blue curve in the right panel of fig.~\ref{figS3}), $\alpha = 1.5$, giving $Z = 19.4$~k$\Omega$. 

In the high-bias regime, the differential conductance is described by,

\begin{equation*}
    \frac{dI}{dV} = \frac{1}{R_{T}} - \frac{2}{\alpha_E R_T} \left( \frac{e}{2 \pi C_T} \right)^2 \frac{1}{V^2}
\end{equation*}

\noindent where $R_T$ is the junction resistance (by fitting T(E) instead of $dI/dV$, $R_T$ serves as a scaling parameter). Using $\alpha_E$ from the fit to the low-bias regime, it is possible to fit the data (red curve in the right panel of fig.~\ref{figS3}) and extract a value of $C_T = 23$~aF. Assuming the tunnel barrier area is defined by the wire width and the well depth, the value of $C_T$ extracted, yields a reasonable tunnel barrier thickness, $t_B = 8.5$~nm.

The black curve in the right panel of fig.~\ref{figS3} shows an approximation of these fitting functions which bridges the two regimes and was used in the analysis in the main paper to extract the varying values of $D_{osc}$.

\section{Alterations to Zhang \emph{et al.}}

N.B. Our axis labelling differs from Zhang -- $z$ and $y$ are interchanged in our treatment.
The Rashba term, $(H_R)_{nn}^{\pm \mp}$ (eqn 22) in our model has a sign change for the different spin species.
We explicitly calculate the expansion coefficients, $a_{n,\sigma}$, from perturbation theory and then calculate the perturbed eigenvalues, $E_{\phi}$, as a function of $k$.
We have reduced the `flattening' field dependence aspect of the unperturbed eigen-energies which we believe to be equivalent to inserting a chemical potential term, $\mu$, to maintain total charge number.

\bibliography{biblio}